\documentclass{emulateapj}
\usepackage{natbib}
\usepackage{epsfig}
\usepackage{graphicx}
\usepackage{subfigure}
\usepackage{float}
\usepackage{amsmath}
\usepackage{color}
\usepackage{amssymb}
\usepackage{amsfonts}
\usepackage{apjfonts}
\usepackage[colorlinks,linkcolor=blue,anchorcolor=green,citecolor=blue]{hyperref}
\shorttitle{Cosmology with GW/FRB Associations}
\shortauthors{Wei et al.}
\begin{document}

\title{Cosmology with Gravitational Wave/Fast Radio Burst Associations}

\author{Jun-Jie Wei,\altaffilmark{1}
Xue-Feng Wu,\altaffilmark{1,2}
and He Gao\altaffilmark{3}}

\affil{$^1$Purple Mountain Observatory, Chinese Academy of Sciences,
Nanjing 210034, China; jjwei@pmo.ac.cn \\
$^2$School of Astronomy and Space Science,
University of Science and Technology of China, Hefei, Anhui 230026, China\\
$^3$Department of Astronomy, Beijing Normal University, Beijing 100875, China}

\begin{abstract}
Recently, some theoretical models predicted that a small fraction of fast radio bursts (FRBs)
could be associated with gravitational waves (GWs). In this work, we discuss the possibility of
using GW/FRB association systems, if commonly detected in the future, as a complementary cosmic probe.
We propose that upgraded standard sirens can be constructed from the joint measurements of
luminosity distances $D_{L}$ derived from GWs and dispersion measures ${\rm DM_{IGM}}$ derived
from FRBs (i.e., the combination $D_{L}\cdot{\rm DM_{IGM}}$). Moreover, unlike the traditional
standard-siren approach (i.e., the $D_{L}$ method) and the $\rm DM_{IGM}$ method that rely on
the optimization of the Hubble constant $H_0$, this $D_{L}\cdot{\rm DM_{IGM}}$ method has
the advantage of being independent of $H_0$. Through Monte Carlo simulations, we prove that
the $D_{L}\cdot{\rm DM_{IGM}}$ method is more effective to constrain cosmological parameters
than $D_{L}$ or ${\rm DM_{IGM}}$ separately, and enables us to achieve accurate multimessenger
cosmology from around 100 GW/FRB systems. Additionally, even if GW/FRB associations are not exist,
the methodology developed here can still be applied to those GWs and FRBs that occur at the same
redshifts.
\end{abstract}


\section{Introduction}
\label{sec:intro}

With the rapid development of modern astronomical technology, the research of cosmology has been promoted into
the age of precision. Cosmological parameters can now be inferred precisely from the observations of various
electromagnetic (EM) waves, such as cosmic microwave background anisotropies \citep{2013ApJS..208...19H,2016A&A...594A..13P},
Type Ia supernovae (\citealt{1998Natur.391...51P,1998AJ....116.1009R}), baryon acoustic oscillations
\citep{2011MNRAS.416.3017B,2012MNRAS.427.3435A}, and so on.

In addition to the traditional EM methods, the observation of gravitational waves (GWs) also provides an alternative probe
for cosmological studies. Due to the fact that the waveform signal of GWs from inspiralling and merging compact binaries
encodes the luminosity distance ($D_{L}$) information, GWs can be considered as standard sirens \citep{1986Natur.323..310S}.
The greatest advantage of GW standard sirens is that the distance calibration is independent of any other cosmic
distance ladders (i.e., it is self-calibrating). Thus, detections of GW together with their EM counterparts
providing the source redshifts, could give the $D_{L}$--$z$ relation for measuring the cosmic expansion
\citep{2005ApJ...629...15H,2011PhRvD..83b3005Z}.
Especially, GW signals from binary neutron stars (NSs) or black hole (BH)--NS mergers are promising for conducting
cosmography, since these merging systems are expected to be accompanied by some detectable EM signals,
e.g., fast radio bursts (FRBs), short Gamma-Ray Bursts (GRBs), or kilonovae/mergernovae
(see \citealt{2016ARNPS..66...23F} for review).
In the past, several works have discussed the possibility of GWs as standard sirens and showed that
with hundreds of simulated GW events they can determine the cosmological parameters with accuracies comparable to
traditional probes (e.g., \citealt{2005ApJ...629...15H,2011PhRvD..83b3005Z,2012PhRvD..86d3011D,2017PhRvD..95d4024C,2017PhRvD..95d3502D}).
Very recently, the coincident detection of a gravitational-wave event GW170817 with EM counterparts (e.g., a GRB 170817A
or a macronova) from a binary NS merger has formally opened the new era of multimessenger astronomy
\citep{2017PhRvL.119p1101A,2017Sci...358.1556C,2017ApJ...848L..14G,2017ApJ...848L..15S}.
Using this first truly GW/EM association, \citet{2017Natur.551...85A} performed a standard siren measurement of the Hubble constant $H_{0}$.

On the other hand, FRBs are a new mysterious class of millisecond-duration radio transients \citep{2007Sci...318..777L,2013Sci...341...53T}.
These objects have anomalously large dispersion measures (DMs), suggesting a cosmological origin for FRBs.
The DM is defined as the integral of the electron number density along the propagation path from the
source to the observer. Since the observed DMs of FRBs contain important information on the cosmological distance
they have traveled, one may combine the DM and $z$ information to probe cosmology if more FRBs with known redshifts can be detected
\citep{2014ApJ...783L..35D,2014ApJ...788..189G,2014ApJ...797...71Z,2014PhRvD..89j7303Z,2016ApJ...830L..31Y}.

Regarding to the physical origins, some studies suggested that mergers of double NSs
\citep{2013PASJ...65L..12T,2016ApJ...822L...7W,2017arXiv171002302Y}, of BH--NS \citep{2015ApJ...814L..20M},
or even of charged BHs \citep{2016ApJ...827L..31Z}, could be responsible for FRBs.
Particularly, \citet{2016ApJ...822L...7W} showed that an FRB could originate from the magnetic interaction
between binary NSs during their final inspiral within the framework of the unipolar inductor model.
The NS--NS merger has been recently confirmed as the progenitor system of GW170817 and GRB 170817A
\citep{2017PhRvL.119p1101A,2017ApJ...848L..14G,2017ApJ...848L..15S}. If FRBs can indeed be
interpreted with the NS--NS merger model, it would be expected to detect possible associations of FRBs
with short GRBs and GW events in the future \citep{2016ApJ...822L...7W}. Alternatively,
\citet{2016ApJ...827L..31Z} proposed that if at least one of the two merging BHs carries
a certain amount of charge, the inspiral process would drive a global magnetic dipole.
The rapid evolution of the magnetic moment of the BH--BH system would lead to a magnetospheric outflow
with an increasing wind power, which may produce an FRB and even a short GRB depending on
the value of the charge. The detection of an FRB associated with future NS--NS (or BH--BH)
merger GW events would verify the NS--NS (or BH--BH) merger model.

In this work, we show that if such GW/FRB association systems are commonly detected in the future,
`upgraded standard sirens' could be constructed from the combination of $D_{L}$ derived from GWs and
DM derived from FRBs, independent of the Hubble constant $H_{0}$. We explore its use to constrain
the cosmological parameters in view of the large samples of GWs and FRBs to be found in
the third-generation GW interferometric detectors such as the Einstein Telescope (ET) and
the upcoming radio transient surveys such as the Square Kilometer Array.

\section{GW/FRB Associations as Upgraded Standard Sirens}
\label{sec:formalism}

\begin{figure*}
\vskip-0.1in
\centerline{\includegraphics[angle=0,width=1.1\hsize]{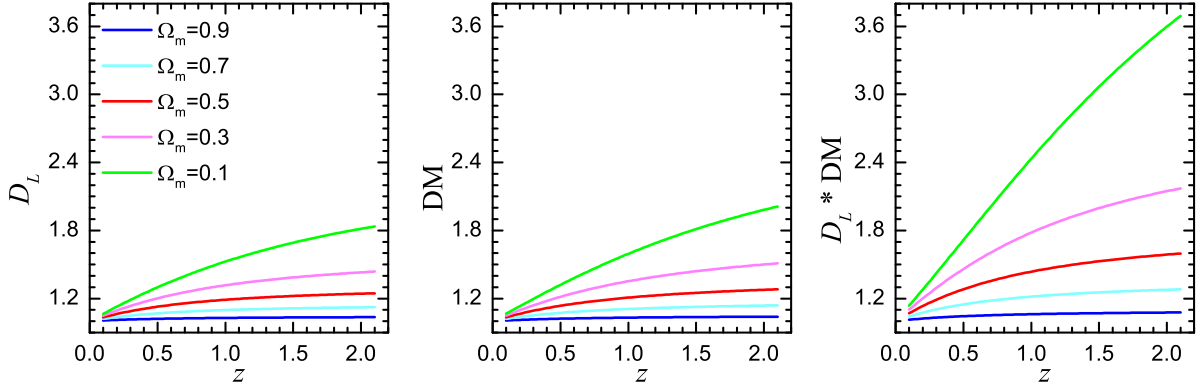}}
\vskip-0.1in
\caption{Sensitivity of three quantities ($D_{L}$, $\rm DM_{IGM}$, and $D_{L}\cdot{\rm DM_{IGM}}$)
to the cosmological parameter. The flat $\Lambda$CDM model is adopted with five different $\Omega_{\rm m}$ values: 0.1, 0.3, 0.5, 0.7, and 0.9.
Each curve is obtained relative to the Einstein-de Sitter Universe.}
\label{fig:f1}
\end{figure*}

\subsection{Luminosity distances from gravitational waves}
The third-generation GW ground-based detectors such as the ET, with ultra high sensitivity,
would significantly improve the detection rate of the GW events. The ET is designed to be ten
times more sensitive than the current advanced laser interferometric detectors, covering the
frequency range of $1-10^{4}$ Hz. It has three interferometers with 10 km arm lengths and
$60^{\circ}$ opening angles, arranged in an equilateral triangle. Here, we present an overview
of using GWs as standard sirens in the potential ET observations (see also \citealt{2017arXiv170300187C}
for a recent review). Throughout we use units $G=c=1$.

The amplitude of the GW depends on the chirp mass and the luminosity distance $D_{L}$.
Because the chirp mass can already be obtained from GW signal's phasing, $D_{L}$ can be extracted
from the amplitude of waveform. In the transverse-traceless gauge, the strain $h(t)$ is
the linear combination of the two components of the GW's tensor (i.e., $h_+$ and $h_\times$),
\begin{equation}
h(t)=F_+(\theta, \phi, \psi)h_+(t)+F_\times(\theta, \phi, \psi)h_\times(t)\;,
\end{equation}
where $F_{+}$ and $F_\times$ are the beam-pattern functions, ($\theta, \phi$) are angles describing
the location of the source relative to the detector, and $\psi$ denotes the polarization angle.
The corresponding antenna pattern functions of one of the interferometers in the ET are~\citep{2011PhRvD..83b3005Z}
\begin{align}
F_+^{(1)}(\theta, \phi, \psi)=&~~\frac{{\sqrt 3 }}{2}[\frac{1}{2}(1 + {\cos ^2}(\theta ))\cos (2\phi )\cos (2\psi ) \nonumber\\
                              &~~- \cos (\theta )\sin (2\phi )\sin (2\psi )],\nonumber\\
F_\times^{(1)}(\theta, \phi, \psi)=&~~\frac{{\sqrt 3 }}{2}[\frac{1}{2}(1 + {\cos ^2}(\theta ))\cos (2\phi )\sin (2\psi ) \nonumber\\
                              &~~+ \cos (\theta )\sin (2\phi )\cos (2\psi )].
\label{equa:F}
\end{align}
The other two interferometers' antenna pattern functions can be derived from Equation~(\ref{equa:F}),
since the interferometers have $60^\circ$ with each other. That is to say,
$F_{+,\times}^{(2)}(\theta, \phi, \psi)=F_{+,\times}^{(1)}(\theta, \phi+2\pi/3, \psi)$ and
$F_{+,\times}^{(3)}(\theta, \phi, \psi)=F_{+,\times}^{(1)}(\theta, \phi+4\pi/3, \psi)$.

In this paper, we focus on the GW signals produced by the merger of binary systems.
Considering a merging binary with component masses $m_1$ and $m_2$, the chirp mass is defined to
be $\mathcal{M}_c=M \eta^{3/5}$, where $M=m_1+m_2$ is the total mass, and $\eta=m_1 m_2/M^2$
represents the symmetric mass ratio. For a GW source locating at cosmological distance with redshift $z$,
the observed chirp mass is given by $\mathcal{M}_{c,\rm obs}=(1+z)\mathcal{M}_{c,\rm phys}$. Below,
$\mathcal{M}_c$ always refers to the observed chirp mass.
Following \citet{2009LRR....12....2S} and \citet{2011PhRvD..83b3005Z}, we apply the stationary phase
approximation to calculate the Fourier transform $\mathcal{H}(f)$ of the time domain waveform $h(t)$,
\begin{equation}
\mathcal{H}(f)=\mathcal{A}f^{-7/6}\exp\left[i\left(2\pi ft_0-\pi/4+2\psi(f/2)-\varphi_{(2,0)}\right)\right]\;,
\label{equa:hf}
\end{equation}
where the constant $t_0$ is the epoch of the merger. The definitions of the functions $\psi$ and $\varphi_{(2,0)}$
are presented in~\citet{2011PhRvD..83b3005Z}. The Fourier amplitude $\mathcal{A}$ is given by
\begin{align}
\mathcal{A}=&~~\frac{1}{D_L}\sqrt{F_+^2\left(1+\cos^2(\iota)\right)^2+4F_\times^2\cos^2(\iota)}\nonumber\\
            &~~\times \sqrt{5\pi/96}\pi^{-7/6}\mathcal{M}_c^{5/6}\;,
\label{equa:A}
\end{align}
where $\iota$ denotes the angle of inclination of the binary's orbital angular momentum with the line-of-sight, and
\begin{equation}
D_{L}(z)=\frac{1+z}{H_0}\int_{0}^{z}\frac{{\rm d}z}{\sqrt{\Omega_{\rm m} (1+z)^{3}+\left(1-\Omega_{\rm m}\right)(1+z)^{3(1+w)}}}
\label{equa:DL}
\end{equation}
is the theoretical luminosity distance in the $w$CDM model. Note that averaging the Fisher matrix over the inclination $\iota$
and the polarization $\psi$ with the constraint $\iota<20^\circ$ is approximately equivalent to taking $\iota=0$.
Therefore, we only consider the simplified case of $\iota=0$ and $\mathcal{A}$ is independent of the polarization angle $\psi$ \citep{2017PhRvD..95d4024C}.

Given the waveform of GWs, we can compute the signal-to-noise ratio (SNR) of the GW detection.
The combined SNR for the network of three independent ET interferometers is
\begin{equation}
\rho=\sqrt{\sum\limits_{i=1}^{3}\left\langle \mathcal{H}^{(i)},\mathcal{H}^{(i)}\right\rangle}\;,
\label{euqa:rho}
\end{equation}
where the inner product is defined as
\begin{equation}
\left\langle{a,b}\right\rangle=4\int_{f_{\rm lower}}^{f_{\rm upper}}\frac{\tilde a(f)\tilde b^\ast(f)+\tilde a^\ast(f)\tilde b(f)}{2}\frac{{\rm d}f}{S_h(f)}\;,
\label{euqa:product}
\end{equation}
where the superscript ``$\thicksim$'' stands for the Fourier transform of the corresponding function
and $S_h(f)$ is the one-side noise power spectral density. We take the ET's $S_h(f)$ to be the same as
in \citet{2011PhRvD..83b3005Z}. The upper cutoff frequency is assumed to be $f_{\rm upper}=2f_{\rm LSO}$,
where $f_{\rm LSO}=1/(6^{3/2}2\pi M_{\rm obs})$ corresponds to the orbit frequency at the last stable orbit,
and $M_{\rm obs}=(1+z)M_{\rm phys}$ is the observed total mass~\citep{2011PhRvD..83b3005Z}. The lower
cutoff frequency $f_{\rm lower}$ is fixed to be $1$ Hz. The signal is identified as a GW event
only when the ET interferometers have a network SNR of $\rho >8.0$.

Using the Fisher information matrix, we can estimate the instrumental uncertainty on the measurement of $D_{L}$,
which can be expressed as ~\citep{2011PhRvD..83b3005Z}
\begin{align}
\sigma_{D_L}^{\rm inst}\simeq \sqrt{\left\langle\frac{\partial \mathcal H}{\partial D_L},\frac{\partial \mathcal H}{\partial D_L}\right\rangle^{-1}}\;.
\end{align}
Assuming that the uncertainty of $D_{L}$ is uncorrelated with the uncertainties of other GW parameters,
we can get $\sigma_{D_L}^{\rm inst}\simeq D_L/\rho$ due to $\mathcal H \propto d_L^{-1}$ \citep{2017PhRvD..95d4024C}.
Taking into account the maximal effect of the inclination $\iota$ on the SNR, we double the estimate of
the error on $D_{L}$, i.e.,
\begin{align}
\sigma_{D_L}^{\rm inst}\simeq \frac{2D_L}{\rho}\;.
\label{sigmainst}
\end{align}
We also add an additional error $\sigma_{D_L}^{\rm lens}/D_L=0.05z$ caused by the weak lensing.
Thus, the total error on $D_{L}$ is given by
\begin{equation}
\sigma_{D_L}=\sqrt{\left(\frac{2D_L}{\rho}\right)^2+\left(0.05z D_L\right)^2}\;.
\label{eq:sigmaDL}
\end{equation}

\subsection{Dispersion measures from fast radio bursts}

In principle, the observed DM of an FRB ($\rm DM_{obs}$; \citealt{2014ApJ...783L..35D,2014ApJ...788..189G,2016ApJ...830L..31Y})
\begin{equation}
{\rm DM_{obs}=DM_{MW}+DM_{IGM}}+\frac{\rm DM_{HG}}{1+z}
\label{eq:DM}
\end{equation}
has contributions from the Milky Way ($\rm DM_{MW}$), intergalactic medium ($\rm DM_{IGM}$), and FRB host galaxy
($\rm DM_{HG}$), respectively. Note that for a GRB-associated FRB, $\rm DM_{HG}$ has contributions from
the host galaxy and the GRB blastwave. Among these terms, $\rm DM_{IGM}$ is the relevant one for cosmological studies.
Considering local inhomogeneity of the intergalactic medium (IGM), we define the average DM of the IGM, which
can be written as \citep{2014ApJ...783L..35D}
\begin{equation}
\left<{\rm DM}_{\rm IGM}\right> = \frac{3H_0\Omega_bf_{\rm IGM}}{8\pi  m_p}
\int_{0}^{z} \frac{\chi(z)(1+z){\rm d} z}{\sqrt{\Omega_{\rm m} (1+z)^{3}+\left(1-\Omega_{\rm m}\right)(1+z)^{3(1+w)}}}\;,
\label{eq:IGM}
\end{equation}
where $f_{\rm IGM}$ is the fraction of baryon mass in the IGM, $\Omega_b$ is the current baryon mass fraction of the universe,
$\chi(z)=(3/4)y_1\chi_{\rm e,H}(z)+(1/8)y_2\chi_{\rm e,He}(z)$, $y_1 \sim 1$ and $y_2 \simeq 4-3y_1 \sim 1$
are the hydrogen (H) and helium (He) mass fractions normalized to $3/4$ and $1/4$, respectively, and $\chi_{\rm e,H}(z)$ and
$\chi_{\rm e,He}(z)$ are the ionization fractions for H and He, respectively. Since H and He are essentially fully ionized
at $z<6$ and at $z<2$ separately \citep{2006ARA&A..44..415F,2009ApJ...694..842M}, it is reasonable to take
$\chi_{\rm e,H}(z)=\chi_{\rm e,He}(z)=1$ for nearby FRBs ($z<2$). One then has $\chi(z)\simeq7/8$.

As long as $\rm DM_{obs}$, $\rm DM_{MW}$, and $\rm DM_{HG}$ can be precisely determined, one can infer the value of
$\left<{\rm DM}_{\rm IGM}\right>$ (see Equation (\ref{eq:DM})).
Then, we can calculate the total uncertainty of $\left<{\rm DM}_{\rm IGM}\right>$ using the expression
\begin{equation}
\sigma_{{\rm DM}_{\rm IGM}} = \left[\sigma_{\rm obs}^{2}+\sigma_{\rm MW}^{2}+\sigma_{\rm IGM}^{2}
+\left(\frac{\sigma_{\rm HG}}{1+z}\right)^{2} \right]^{1/2}\;.
\label{eq:sigmaIGM}
\end{equation}
Following \citet{2014ApJ...788..189G}, we investigate different contributions of the relevant uncertainties in Equation~(\ref{eq:sigmaIGM}) below.

Up to FRB 180311, a total of 33 FRBs have been detected \citep{2016PASA...33...45P}. The measurements of $\rm DM_{obs}$
and the corresponding uncertainties for these 33 FRBs are available in the FRB catalogue\footnote{http://frbcat.org/}.
Here, we adopt an average of these values as the uncertainty of $\rm DM_{obs}$, i.e., $\sigma_{\rm obs}=1.5$ pc $\rm cm^{-3}$.
With the ATNF pulsar catalogue \citep{2005AJ....129.1993M}\footnote{http://www.atnf.csiro.au/research/pulsar/psrcat/},
we find that the average uncertainty of $\rm DM_{MW}$ for high Galactic latitude ($|b|>10^{\circ}$) sources is about
$10$ pc $\rm cm^{-3}$, and we adopt this value as $\sigma_{\rm MW}$. To be conservative, we associate an uncertainty
of $\sigma_{\rm IGM}=100$ pc $\rm cm^{-3}$ to $\rm DM_{IGM}$, as \citet{2016ApJ...830L..31Y} did in their treatment,
with the hope that such a large uncertainty could account for the IGM inhomogeneity effect. On the basis of the DM uncertainty
of the Milky Way, one may deduce that the uncertainty of $\rm DM_{HG}$ could be from tens to hundreds of pc $\rm cm^{-3}$.
In addition, \citet{2014ApJ...788..189G} showed that the resulting constraints on cosmological parameters are not
very sensitive to the value of $\sigma_{\rm HG}$, since $\sigma_{\rm HG}$ becomes less significant at high redshifts
due to the $(1+z)$ factor. Here we adopt $\sigma_{\rm HG}=30$ pc $\rm cm^{-3}$.

\subsection{The combination of $D_{L}$ and DM}
If FRBs are confirmed to be associated with GW events, the combination of $D_{L}$ measurements of GWs and
DM measurements of FRBs could provide upgraded standard sirens to study cosmology. From Equations~(\ref{equa:DL})
and~(\ref{eq:IGM}), we can see that the Hubble constant $H_0$ cancels out when we multiply $D_{L}$ by $\rm DM_{IGM}$,
so the constraints on the cosmological parameters from the product $D_{L}\cdot{\rm DM_{IGM}}$ are independent of
the Hubble constant. With the combination of $D_{L}\cdot{\rm DM_{IGM}}$, the propagated error $\sigma_{D_{L}\cdot{\rm DM}}$
in $D_{L}\cdot{\rm DM_{IGM}}$ is
\begin{equation}
\sigma_{D_{L}\cdot{\rm DM}} = \left[\left({\rm DM_{IGM}}\cdot\sigma_{D_{L}}\right)^{2}+\left(D_{L}\cdot\sigma_{{\rm DM_{IGM}}}\right)^{2}\right]^{1/2}\;.
\end{equation}

In Figure~\ref{fig:f1}, we illustrate the three quantities ($D_{L}$, $\rm DM_{IGM}$, and $D_{L}\cdot{\rm DM_{IGM}}$)
as a function of the redshift $z$ in the flat $\Lambda$CDM model. To show the sensitivity of the three functions to
the cosmological parameter $\Omega_{\rm m}$, we plot them for five cases of a flat Universe with
$\Omega_{\rm m}=0.1,\;0.3,\;0.5,\;0.7$, and $0.9$, relative to an Einstein-de Sitter Universe ($\Omega_{\rm m}=1,\;\Omega_{\Lambda}=0$).
It is clearly seen that the $D_{L}\cdot{\rm DM_{IGM}}$ curves have a wider separation than the $D_{L}$ or $\rm DM_{IGM}$ curves
to allow a better discrimination among different cosmological models. Meanwhile, the sensitivity increases with
the redshift, thus, it is of special significance for the $D_{L}\cdot{\rm DM_{IGM}}$ method to study high-redshift associations.

\subsection{Redshifts from electromagnetic counterparts}
Measuring the source redshift is crucial when using the GW/FRB association as the upgraded standard siren.
Several methods have been suggested to obtain the redshift associated to a GW event, such as the galaxy catalogue
\citep{1986Natur.323..310S}, NS mass distribution \citep{1993PhRvD..48.4738M,2012PhRvD..85b3535T}, and
the tidal deformation of NSs \citep{2012PhRvL.108i1101M}. In this work, we adopt the widely used method of
identifying an EM counterpart of the GW event to obtain the source redshift \citep{2010ApJ...725..496N,2010CQGra..27u5006S,2011PhRvD..83b3005Z}.
An EM counterpart like the GRB or the kilonova can give the redshift information if the host galaxy of the event
can be pinpointed. Besides, the redshift can also be measured from the absorption lines of the GRB afterglows.

\section{Monte Carlo Simulations}
\label{sec:MC}
To explore the cosmological constraint ability by future joint measurements of luminosity distance $D_{L}$
and dispersion measure DM, we perform Monte Carlo simulations on GW/FRB systems. Here we adopt the cosmological
parameters of the fiducial flat $\Lambda$CDM model derived from \emph{Planck} 2015 data: $H_{0}=67.8$ km $\rm s^{-1}$ $\rm Mpc^{-1}$,
$\Omega_{\rm m}=0.308$, $\Omega_{\Lambda}=0.692$, and $\Omega_b=0.049$ \citep{2016A&A...594A..13P}.
For the fraction of baryon mass in the IGM, we take $f_{\rm IGM}=0.83$ \citep{1998ApJ...503..518F,2012ApJ...759...23S,2014ApJ...783L..35D}.
If compact binaries are NS--NS
binaries or NS--BH binaries, it is believed that the source redshift can be obtained from an EM counterpart that
occurs coincidentally with the GW event \citep{2010ApJ...725..496N,2010CQGra..27u5006S,2011PhRvD..83b3005Z}.
Moreover, \citet{2016ApJ...822L...7W} proposed that possible GW/FRB associations could be detected within
the framework of the NS--NS merger model. Therefore, we consider the mergers of binary NS systems as the
sources of GWs and FRBs. Following \citet{2011PhRvD..83b3005Z} and \citet{2017PhRvD..95d4024C},
the redshift distribution of the sources takes the form
\begin{equation}
P(z)\propto \frac{4\pi D_C^2(z)R(z)}{H(z)(1+z)},
\label{equa:pz}
\end{equation}
where $D_C(z)=\int_0^z {1/H(z){\rm d}z}$ is the comoving distance, and $R(z)$ denotes the time evolution of
the merger rate and takes the form~\citep{2001MNRAS.324..797S,2009PhRvD..80j4009C,2017PhRvD..95d4024C}
\begin{equation}
R(z)=\begin{cases}
1+2z, & z\leq 1 \\
\frac{3}{4}(5-z), & 1<z<5 \\
0, & z\geq 5.
\end{cases}
\label{equa:rz}
\end{equation}
In our simulations, the redshifts of source $z$ are randomly generated from the redshift probability distribution function
(Equation~(\ref{equa:pz})). Since the ET would be able to detect binary NS inspirals up to redshifts of $z\sim2$,
the range of the source redshift $z$ for our analysis is from $0$ to $2$. With the mock $z$, we infer the fiducial values
of $D_{L}^{\rm fid}$ and $\rm DM_{IGM}^{fid}$ from Equations~(\ref{equa:DL}) and (\ref{eq:IGM}), respectively.
The mass of each NS and the position angle $\theta$ are uniformly distributed in the two parameter intervals:
$[1,2]$ $M_{\odot}$ and $[0,\pi]$, respectively\footnote{We do not need to consider the other two angles $\phi$ and $\psi$,
since the SNR is independent of them.}. We then calculate the combined SNR of each set of the random sample using Equation~(\ref{euqa:rho}),
and confirm that the simulated signal is a GW detection if $\rho >8.0$. For every confirmed detection,
we add the deviations in Equations~(\ref{eq:sigmaDL}) and (\ref{eq:sigmaIGM}) to the fiducial values of $D_{L}^{\rm fid}$ and $\rm DM_{IGM}^{fid}$,
respectively. That is, we sample the $D_{L}^{\rm mea}$ (or $\rm DM_{IGM}^{mea}$) measurement according to the Gaussian distribution
$D_{L}^{\rm mea}=\mathcal{N}(D_{L}^{\rm fid},\;\sigma_{D_{L}})$ (or ${\rm DM_{IGM}^{mea}}=\mathcal{N}({\rm DM_{IGM}^{fid}},\;\sigma_{{\rm DM}_{\rm IGM}})$).
The inferred event rate density of NS--NS mergers from the detection of GW170817 is $\sim1100^{+2500}_{-910}$ $\rm Gpc^{-3}$ $\rm yr^{-1}$
\citep{2017PhRvL.119p1101A}. The event rate density of FRBs may be estimated as \citep{2016ApJ...827L..31Z}
$\dot{\rho}_{\rm FRB}=\frac{365\dot{N}_{\rm FRB}}{\left(4\pi/3\right)D_{z}^{3}}\simeq\left(1.4\times10^{3}\;{\rm Gpc^{-3}\;yr^{-1}}\right)
\left(\frac{D_{z}}{\rm 3.4\; Gpc}\right)^{-3}\left(\frac{\dot{N}_{\rm FRB}}{10^{4}}\right)$, where $D_{z}$ is the comoving distance
of the FRB normalized to 3.4 Gpc ($z=1$), and $\dot{N}_{\rm FRB}$ denotes the daily all-sky FRB rate that is normalized to $10^{4}$.
One can see that the FRB rate is well consistent with the NS--NS merger rate.
The expected detection rates of NS--NS and BH--NS per year for the ET\footnote{The Einstein Telescope Project, https://www.et.gw.eu/et/.}
are about the order $10^{3}-10^{7}$. Taking the detection rate in the middle rang $\mathcal{O}(10^5)$, and assuming that
only a small fraction ($\sim 10^{-3}$) of GW/FRB systems could be detected, we can expect to detect $\mathcal{O}(10^2)$ such systems per year.
Thus, we simulate a population of 100 GW/FRB systems.

\begin{figure}
\vskip-0.1in
\centerline{\includegraphics[angle=0,width=0.48\textwidth]{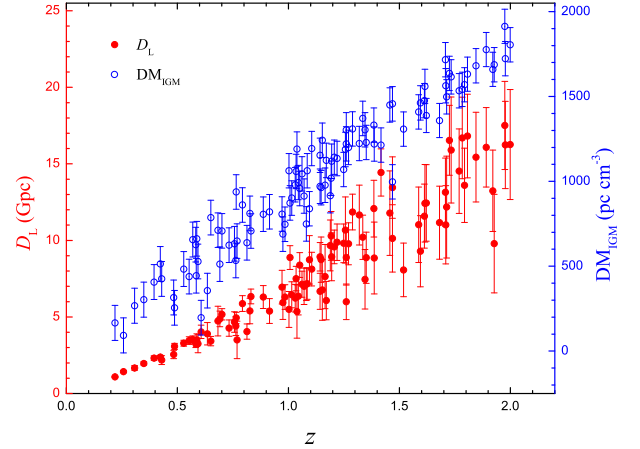}}
\vskip-0.1in
\caption{An example catalogue of 100 simulated GW/FRB associations with redshifts $z$,
luminosity distances $D_{L}$, and dispersion measures $\rm DM_{IGM}$.}
\label{fig:f2}
\vskip 0.1in
\end{figure}

An example of 100 simulated GW/FRB systems from the fiducial model is shown in Figure~\ref{fig:f2}.
For a set of 100 simulated data points, the likelihood for the cosmological parameters can be determined
by the minimum $\chi^{2}$ statistic, i.e.,
\begin{equation}
\chi^2(\mathbf{p})=\sum_{i}\frac{\left[D_{L}^{\rm mea}\cdot{\rm DM_{IGM}^{mea}}-D_{L}^{\rm th}(\mathbf{p})\cdot{\rm DM_{IGM}^{th}}(\mathbf{p})\right]^2}{\sigma^{2}_{D_{L}\cdot{\rm DM}}}\;,
\end{equation}
where $D_{L}^{\rm th}\cdot{\rm DM_{IGM}^{th}}$ is the theoretical value calculated from the set of cosmological parameters $\mathbf{p}$.
Similar expressions are computed for the methods of $D_{L}$ and $\rm DM_{IGM}$. This simulation is repeated for 1000 times
to ensure the final constraint results are unbiased.

\begin{figure}
\vskip-0.5in
\centerline{\includegraphics[angle=0,width=0.35\textwidth]{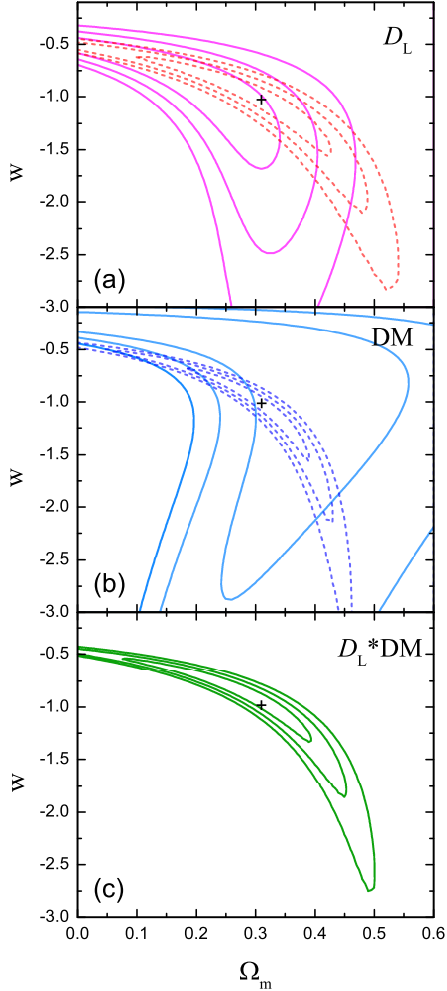}}
\vskip-0.6in
\caption{Constraint results of $(\Omega_{\rm m},\;w)$ in the $w$CDM model from 100 simulated GW/FRB systems using three different methods (from top to bottom):
$D_{L}$, $\rm DM_{IGM}$, and $D_{L}\cdot{\rm DM_{IGM}}$. Unlike the methods of $D_{L}$ and $\rm DM_{IGM}$ that
rely on the optimization of $H_{0}$, the $D_{L}\cdot{\rm DM_{IGM}}$ method is independent of $H_{0}$.
The dashed and solid contours in panels (a) and (b) correspond to the cases of fixing and marginalizing over $H_{0}$, respectively.
The plus symbols denote the simulated values.}
\label{fig:f3}
\vskip 0.1in
\end{figure}

In $w$CDM, the equation-of-state of dark energy, $w$, is constant, and there are three free parameters:
$\Omega_{\rm m}$, $w$, and $H_{0}$. It should be underlined that the $D_{L}\cdot{\rm DM_{IGM}}$ method
(the product of Equations~(\ref{equa:DL}) and (\ref{eq:IGM}))
can be used to test cosmological models in a rather unique way because, unlike the other two methods
($D_{L}$ or $\rm DM_{IGM}$) that rely on the optimization of the Hubble constant $H_0$, this particular
analysis is completely independent of $H_0$. For the $D_{L}$ and $\rm DM_{IGM}$ methods, we let
$\Omega_{\rm m}$ and $w$ to be free parameters while either fixing or marginalizing over $H_0$.

We first marginalize $H_0$ in the $w$CDM model to find the confidence levels in the $\Omega_{\rm m}-w$ plane.
The constraint results (solid lines) from three different methods ($D_{L}$, $\rm DM_{IGM}$, and $D_{L}\cdot{\rm DM_{IGM}}$)
are illustrated in Figure~\ref{fig:f3}. One can see from these solid contours that the $D_{L}\cdot{\rm DM_{IGM}}$ method
gives much tighter constraints on both cosmological parameters than the other two methods as we expected.
In both the traditional standard-siren approach (i.e., the $D_{L}$ method) and the $\rm DM_{IGM}$ method,
we need a much larger sample to increase the significance of the constraints. In contrast, future observations
of GWs and their FRB counterparts will enable us to achieve precise cosmography from around 100 such systems.
All in all, upgraded standard sirens could be constructed if GW/FRB association systems are commonly detected
in the future.

To show the importance of $H_0$ in the $D_{L}$ and $\rm DM_{IGM}$ methods, we also present the case of fixing
$H_{0}=67.8$ km $\rm s^{-1}$ $\rm Mpc^{-1}$. As shown by the dashed contours in Figures~\ref{fig:f3}(a) and (b),
the constraints on cosmological parameters can be significantly improved when $H_{0}$ is fixed. Even if the prior
value of $H_{0}$ is adopted, however, the constraints obtained from the methods of $D_{L}$ and $\rm DM_{IGM}$
are still not better than that of the $D_{L}\cdot{\rm DM_{IGM}}$ method (Figure~\ref{fig:f3}c). Therefore, we can
conclude that the cosmological constraint ability of the methods of $D_{L}$ and $\rm DM_{IGM}$ are restricted
by the fact that they both explicitly depend on $H_{0}$, while the $D_{L}\cdot{\rm DM_{IGM}}$ method has the
advantage of being independent of $H_{0}$.

\begin{figure}
\vskip-0.1in
\centerline{\includegraphics[angle=0,width=0.5\textwidth]{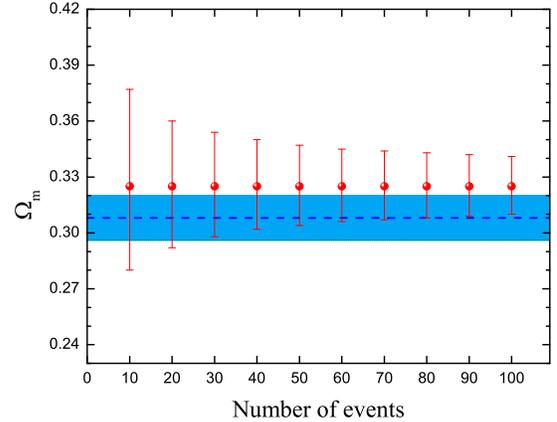}}
\vskip-0.1in
\caption{The best-fit $\Omega_{\rm m}$ (red dot) and $1\sigma$ confidence level (red line)
in the flat $\Lambda$CDM model as a function of the number of GW/FRB associations.
The dashed line is the fiducial value. The blue shaded area represents the $1\sigma$ confidence level
constraint from Planck 2015 results.}
\label{fig:f4}
\vskip 0.1in
\end{figure}

To better represent how effective this $D_{L}\cdot{\rm DM_{IGM}}$ method might be with a certain number
of coincident detections, in Figure~\ref{fig:f4} we plot the best-fit dark matter density parameter
$\Omega_{\rm m}$ and $1\sigma$ confidence level in the flat $\Lambda$CDM model as a function of the number of
GW/FRB associations (analogous to Figure~5 of \citet{2012PhRvD..86d3011D}). The $1\sigma$ confidence level
constraint on $\Omega_{\rm m}$ from Planck temperature data and Planck lensing combined results (blue shaded area;
\citealt{2016A&A...594A..13P}) is also plotted for comparison. One can see from this figure that with about
100 GW/FRB associations we can constrain $\Omega_{\rm m}$ with an accuracy comparable to Planck data.

\section{Summary and Discussion}
\label{sec:summary}
In this work, we propose that if GW/FRB associations are confirmed to commonly exist,
upgraded standard sirens can be constructed from the joint measurements of luminosity
distances $D_{L}$ derived from GWs and dispersion measures ${\rm DM_{IGM}}$ derived from FRBs.
Moreover, the combination of $D_{L}$ and ${\rm DM_{IGM}}$ (i.e., the $D_{L}\cdot{\rm DM_{IGM}}$ method)
can be used to differentiate cosmological models in a rather unique way because, unlike the traditional
standard-siren approach (i.e., the $D_{L}$ method) and the $\rm DM_{IGM}$ method that rely on
the optimization of the Hubble constant $H_0$, this particular analysis is completely independent of $H_0$.
Through Monte Carlo simulations, we prove that this $D_{L}\cdot{\rm DM_{IGM}}$ method
is able to constrain the cosmological parameters more strongly than $D_{L}$ or ${\rm DM_{IGM}}$ separately.
With the help of the $D_{L}\cdot{\rm DM_{IGM}}$ method, precise multimessenger cosmology can be achieved from around 100 GW/FRB systems.

Thanks to the high sensitivity, the planned third-generation GW detectors, such as the ET,
could detect about $10^{3}-10^{7}$ NS--NS and BH-NS merger GW events per year. Although
a considerable catalogue of GW events would be obtained, the measurements of GW/FRB association
systems suggested by our method may not be easy in practice. To be specific, only in the optimistic
case that satisfying (i) GWs and FRBs are confirmed to have the same progenitor system; (ii)
the coincident detections of GW events with FRBs can be accomplished by the collaboration of
the GW interferometric detectors and the radio transient surveys; (iii) the source redshifts
can be identified; (iv) an overall statistical error from the contribution of $\rm DM_{obs}$,
$\rm DM_{MW}$, and $\rm DM_{HG}$ is smaller than the systematic uncertainty $\sigma_{\rm IGM}$ in modeling and
inferring the dispersion measure $\rm DM_{IGM}$ of the intergalactic plasma,\footnote{If it
turns out that the statistical error could become larger than $\sigma_{\rm IGM}$, the accuracy
of our results would be reduced dramatically.} and $\sigma_{\rm IGM}$ is not too big, GW/FRB systems could serve as a viable cosmic probe.
If some of these requirements are not met, the using of GW/FRB systems as upgraded standard sirens
would be challenged. Although in this work we just discuss FRBs, the methodology developed here is also applicable for
any other kinds of cosmological radio transients that occur simultaneously with GWs, if there are,
to constrain the cosmological parameters and the equation of state of dark energy with high accuracy.
On the other hand, even if FRBs are not associated with GWs, our method is still applicable for those
GWs and FRBs that occur at the same redshifts.

\acknowledgments
We thank the anonymous referee for constructive suggestions.
This work is partially supported by the National Basic Research Program (``973'' Program)
of China (Grant No. 2014CB845800), the National Natural Science Foundation of China
(Grant Nos. 11603076, 11673068, and 11725314), the Youth Innovation Promotion
Association (2011231 and 2017366), the Key Research Program of Frontier Sciences (Grant No. QYZDB-SSW-SYS005),
the Strategic Priority Research Program ``Multi-waveband gravitational wave Universe''
(Grant No. XDB23000000) of the Chinese Academy of Sciences, and the Natural Science Foundation
of Jiangsu Province (Grant No. BK20161096).

\bibliographystyle{apj}

\begin{thebibliography}{}
\expandafter\ifx\csname natexlab\endcsname\relax\def\natexlab#1{#1}\fi

\bibitem[{{Abbott} {et~al.}(2017{\natexlab{a}}){Abbott}, {Abbott}, {Abbott},
  {Acernese}, {Ackley}, {Adams}, {Adams}, {Addesso}, {Adhikari}, {Adya}, \&
  et~al.}]{2017PhRvL.119p1101A}
{Abbott}, B.~P., {Abbott}, R., {Abbott}, T.~D., {et~al.} 2017{\natexlab{a}}, Physical Review Letters, 119, 161101

\bibitem[{{Abbott} {et~al.}(2017{\natexlab{b}}){Abbott}, {Abbott}, {Abbott},
  {Acernese}, {Ackley}, {Adams}, {Adams}, {Addesso}, {Adhikari}, {Adya}, \&
  et~al.}]{2017Natur.551...85A}
{Abbott}, B.~P., {Abbott}, R., {Abbott}, T.~D., {et~al.} 2017{\natexlab{b}},
  \nat, 551, 85

\bibitem[{{Anderson} {et~al.}(2012){Anderson}, {Aubourg}, {Bailey}, {Bizyaev},
  {Blanton}, {Bolton}, {Brinkmann}, {Brownstein}, {Burden}, {Cuesta}, {da
  Costa}, {Dawson}, {de Putter}, {Eisenstein}, {Gunn}, {Guo}, {Hamilton},
  {Harding}, {Ho}, {Honscheid}, {Kazin}, {Kirkby}, {Kneib}, {Labatie},
  {Loomis}, {Lupton}, {Malanushenko}, {Malanushenko}, {Mandelbaum}, {Manera},
  {Maraston}, {McBride}, {Mehta}, {Mena}, {Montesano}, {Muna}, {Nichol},
  {Nuza}, {Olmstead}, {Oravetz}, {Padmanabhan}, {Palanque-Delabrouille}, {Pan},
  {Parejko}, {P{\^a}ris}, {Percival}, {Petitjean}, {Prada}, {Reid}, {Roe},
  {Ross}, {Ross}, {Samushia}, {S{\'a}nchez}, {Schlegel}, {Schneider},
  {Sc{\'o}ccola}, {Seo}, {Sheldon}, {Simmons}, {Skibba}, {Strauss}, {Swanson},
  {Thomas}, {Tinker}, {Tojeiro}, {Maga{\~n}a}, {Verde}, {Wagner}, {Wake},
  {Weaver}, {Weinberg}, {White}, {Xu}, {Y{\`e}che}, {Zehavi}, \&
  {Zhao}}]{2012MNRAS.427.3435A}
{Anderson}, L., {Aubourg}, E., {Bailey}, S., {et~al.} 2012, \mnras, 427, 3435

\bibitem[{{Beutler} {et~al.}(2011){Beutler}, {Blake}, {Colless}, {Jones},
  {Staveley-Smith}, {Campbell}, {Parker}, {Saunders}, \&
  {Watson}}]{2011MNRAS.416.3017B}
{Beutler}, F., {Blake}, C., {Colless}, M., {et~al.} 2011, \mnras, 416, 3017

\bibitem[{{Cai} {et~al.}(2017){Cai}, {Cao}, {Guo}, {Wang}, \&
  {Yang}}]{2017arXiv170300187C}
{Cai}, R.-G., {Cao}, Z., {Guo}, Z.-K., {Wang}, S.-J., \& {Yang}, T. 2017, Natl.
  Sci. Rev., 4, 687

\bibitem[{{Cai} \& {Yang}(2017)}]{2017PhRvD..95d4024C}
{Cai}, R.-G., \& {Yang}, T. 2017, \prd, 95, 044024


\bibitem[{{Coulter} {et~al.}(2017){Coulter}, {Foley}, {Kilpatrick}, {Drout},
  {Piro}, {Shappee}, {Siebert}, {Simon}, {Ulloa}, {Kasen}, {Madore},
  {Murguia-Berthier}, {Pan}, {Prochaska}, {Ramirez-Ruiz}, {Rest}, \&
  {Rojas-Bravo}}]{2017Sci...358.1556C}
{Coulter}, D.~A., {Foley}, R.~J., {Kilpatrick}, C.~D., {et~al.} 2017, Science,
  358, 1556

\bibitem[{{Cutler} \& {Holz}(2009)}]{2009PhRvD..80j4009C}
{Cutler}, C., \& {Holz}, D.~E. 2009, \prd, 80, 104009

\bibitem[{{Del Pozzo}(2012)}]{2012PhRvD..86d3011D}
{Del Pozzo}, W. 2012, \prd, 86, 043011

\bibitem[{{Del Pozzo} {et~al.}(2017){Del Pozzo}, {Li}, \&
  {Messenger}}]{2017PhRvD..95d3502D}
{Del Pozzo}, W., {Li}, T.~G.~F., \& {Messenger}, C. 2017, \prd, 95, 043502

\bibitem[{{Deng} \& {Zhang}(2014)}]{2014ApJ...783L..35D}
{Deng}, W., \& {Zhang}, B. 2014, \apjl, 783, L35

\bibitem[{{Fan} {et~al.}(2006){Fan}, {Carilli}, \&
  {Keating}}]{2006ARA&A..44..415F}
{Fan}, X., {Carilli}, C.~L., \& {Keating}, B. 2006, ARA\&A, 44, 415

\bibitem[{{Fern{\'a}ndez} \& {Metzger}(2016)}]{2016ARNPS..66...23F}
{Fern{\'a}ndez}, R., \& {Metzger}, B.~D. 2016, Annual Review of Nuclear and
  Particle Science, 66, 23

\bibitem[{{Fukugita} {et~al.}(1998){Fukugita}, {Hogan}, \&
  {Peebles}}]{1998ApJ...503..518F}
{Fukugita}, M., {Hogan}, C.~J., \& {Peebles}, P.~J.~E. 1998, \apj, 503, 518

\bibitem[{{Gao} {et~al.}(2014){Gao}, {Li}, \& {Zhang}}]{2014ApJ...788..189G}
{Gao}, H., {Li}, Z., \& {Zhang}, B. 2014, \apj, 788, 189

\bibitem[{{Goldstein} {et~al.}(2017){Goldstein}, {Veres}, {Burns}, {Briggs},
  {Hamburg}, {Kocevski}, {Wilson-Hodge}, {Preece}, {Poolakkil}, {Roberts},
  {Hui}, {Connaughton}, {Racusin}, {von Kienlin}, {Dal Canton}, {Christensen},
  {Littenberg}, {Siellez}, {Blackburn}, {Broida}, {Bissaldi}, {Cleveland},
  {Gibby}, {Giles}, {Kippen}, {McBreen}, {McEnery}, {Meegan}, {Paciesas}, \&
  {Stanbro}}]{2017ApJ...848L..14G}
{Goldstein}, A., {Veres}, P., {Burns}, E., {et~al.} 2017, \apjl, 848, L14

\bibitem[{{Hinshaw} {et~al.}(2013){Hinshaw}, {Larson}, {Komatsu}, {Spergel},
  {Bennett}, {Dunkley}, {Nolta}, {Halpern}, {Hill}, {Odegard}, {Page}, {Smith},
  {Weiland}, {Gold}, {Jarosik}, {Kogut}, {Limon}, {Meyer}, {Tucker}, {Wollack},
  \& {Wright}}]{2013ApJS..208...19H}
{Hinshaw}, G., {Larson}, D., {Komatsu}, E., {et~al.} 2013, \apjs, 208, 19

\bibitem[{{Holz} \& {Hughes}(2005)}]{2005ApJ...629...15H}
{Holz}, D.~E., \& {Hughes}, S.~A. 2005, \apj, 629, 15

\bibitem[{{Lorimer} {et~al.}(2007){Lorimer}, {Bailes}, {McLaughlin},
  {Narkevic}, \& {Crawford}}]{2007Sci...318..777L}
{Lorimer}, D.~R., {Bailes}, M., {McLaughlin}, M.~A., {Narkevic}, D.~J., \&
  {Crawford}, F. 2007, Science, 318, 777

\bibitem[{{Manchester} {et~al.}(2005){Manchester}, {Hobbs}, {Teoh}, \&
  {Hobbs}}]{2005AJ....129.1993M}
{Manchester}, R.~N., {Hobbs}, G.~B., {Teoh}, A., \& {Hobbs}, M. 2005, \aj, 129,
  1993

\bibitem[{{Markovi{\'c}}(1993)}]{1993PhRvD..48.4738M}
{Markovi{\'c}}, D. 1993, \prd, 48, 4738

\bibitem[{{McQuinn} {et~al.}(2009){McQuinn}, {Lidz}, {Zaldarriaga},
  {Hernquist}, {Hopkins}, {Dutta}, \&
  {Faucher-Gigu{\`e}re}}]{2009ApJ...694..842M}
{McQuinn}, M., {Lidz}, A., {Zaldarriaga}, M., {et~al.} 2009, \apj, 694, 842

\bibitem[{{Messenger} \& {Read}(2012)}]{2012PhRvL.108i1101M}
{Messenger}, C., \& {Read}, J. 2012, Physical Review Letters, 108, 091101

\bibitem[{{Mingarelli} {et~al.}(2015){Mingarelli}, {Levin}, \&
  {Lazio}}]{2015ApJ...814L..20M}
{Mingarelli}, C.~M.~F., {Levin}, J., \& {Lazio}, T.~J.~W. 2015, \apjl, 814, L20

\bibitem[{{Nissanke} {et~al.}(2010){Nissanke}, {Holz}, {Hughes}, {Dalal}, \&
  {Sievers}}]{2010ApJ...725..496N}
{Nissanke}, S., {Holz}, D.~E., {Hughes}, S.~A., {Dalal}, N., \& {Sievers},
  J.~L. 2010, \apj, 725, 496

\bibitem[{{Perlmutter} {et~al.}(1998){Perlmutter}, {Aldering}, {della Valle},
  {Deustua}, {Ellis}, {Fabbro}, {Fruchter}, {Goldhaber}, {Groom}, {Hook},
  {Kim}, {Kim}, {Knop}, {Lidman}, {McMahon}, {Nugent}, {Pain}, {Panagia},
  {Pennypacker}, {Ruiz-Lapuente}, {Schaefer}, \&
  {Walton}}]{1998Natur.391...51P}
{Perlmutter}, S., {Aldering}, G., {della Valle}, M., {et~al.} 1998, \nat, 391,
  51

\bibitem[{{Petroff} {et~al.}(2016){Petroff}, {Barr}, {Jameson}, {Keane},
  {Bailes}, {Kramer}, {Morello}, {Tabbara}, \& {van
  Straten}}]{2016PASA...33...45P}
{Petroff}, E., {Barr}, E.~D., {Jameson}, A., {et~al.} 2016, PASA, 33, e045

\bibitem[{{Planck Collaboration} {et~al.}(2016){Planck Collaboration}, {Ade},
  {Aghanim}, {Arnaud}, {Ashdown}, {Aumont}, {Baccigalupi}, {Banday},
  {Barreiro}, {Bartlett}, \& et~al.}]{2016A&A...594A..13P}
{Planck Collaboration}, {Ade}, P.~A.~R., {Aghanim}, N., {et~al.} 2016, \aap,
  594, A13

\bibitem[{{Riess} {et~al.}(1998){Riess}, {Filippenko}, {Challis},
  {Clocchiatti}, {Diercks}, {Garnavich}, {Gilliland}, {Hogan}, {Jha},
  {Kirshner}, {Leibundgut}, {Phillips}, {Reiss}, {Schmidt}, {Schommer},
  {Smith}, {Spyromilio}, {Stubbs}, {Suntzeff}, \&
  {Tonry}}]{1998AJ....116.1009R}
{Riess}, A.~G., {Filippenko}, A.~V., {Challis}, P., {et~al.} 1998, \aj, 116,
  1009

\bibitem[{{Sathyaprakash} \& {Schutz}(2009)}]{2009LRR....12....2S}
{Sathyaprakash}, B.~S., \& {Schutz}, B.~F. 2009, Living Reviews in Relativity,
  12, 2

\bibitem[{{Sathyaprakash} {et~al.}(2010){Sathyaprakash}, {Schutz}, \& {Van Den
  Broeck}}]{2010CQGra..27u5006S}
{Sathyaprakash}, B.~S., {Schutz}, B.~F., \& {Van Den Broeck}, C. 2010,
  Classical and Quantum Gravity, 27, 215006

\bibitem[{{Savchenko} {et~al.}(2017){Savchenko}, {Ferrigno}, {Kuulkers},
  {Bazzano}, {Bozzo}, {Brandt}, {Chenevez}, {Courvoisier}, {Diehl}, {Domingo},
  {Hanlon}, {Jourdain}, {von Kienlin}, {Laurent}, {Lebrun}, {Lutovinov},
  {Martin-Carrillo}, {Mereghetti}, {Natalucci}, {Rodi}, {Roques}, {Sunyaev}, \&
  {Ubertini}}]{2017ApJ...848L..15S}
{Savchenko}, V., {Ferrigno}, C., {Kuulkers}, E., {et~al.} 2017, \apjl, 848, L15

\bibitem[{{Schneider} {et~al.}(2001){Schneider}, {Ferrari}, {Matarrese}, \&
  {Portegies Zwart}}]{2001MNRAS.324..797S}
{Schneider}, R., {Ferrari}, V., {Matarrese}, S., \& {Portegies Zwart}, S.~F.
  2001, \mnras, 324, 797

\bibitem[{{Schutz}(1986)}]{1986Natur.323..310S}
{Schutz}, B.~F. 1986, \nat, 323, 310

\bibitem[{{Shull} {et~al.}(2012){Shull}, {Smith}, \&
  {Danforth}}]{2012ApJ...759...23S}
{Shull}, J.~M., {Smith}, B.~D., \& {Danforth}, C.~W. 2012, \apj, 759, 23

\bibitem[{{Taylor} {et~al.}(2012){Taylor}, {Gair}, \&
  {Mandel}}]{2012PhRvD..85b3535T}
{Taylor}, S.~R., {Gair}, J.~R., \& {Mandel}, I. 2012, \prd, 85, 023535

\bibitem[{{Thornton} {et~al.}(2013){Thornton}, {Stappers}, {Bailes},
  {Barsdell}, {Bates}, {Bhat}, {Burgay}, {Burke-Spolaor}, {Champion}, {Coster},
  {D'Amico}, {Jameson}, {Johnston}, {Keith}, {Kramer}, {Levin}, {Milia}, {Ng},
  {Possenti}, \& {van Straten}}]{2013Sci...341...53T}
{Thornton}, D., {Stappers}, B., {Bailes}, M., {et~al.} 2013, Science, 341, 53

\bibitem[{{Totani}(2013)}]{2013PASJ...65L..12T}
{Totani}, T. 2013, \pasj, 65, L12

\bibitem[{{Wang} {et~al.}(2016){Wang}, {Yang}, {Wu}, {Dai}, \&
  {Wang}}]{2016ApJ...822L...7W}
{Wang}, J.-S., {Yang}, Y.-P., {Wu}, X.-F., {Dai}, Z.-G., \& {Wang}, F.-Y. 2016,
  \apjl, 822, L7

\bibitem[{{Yamasaki} {et~al.}(2017){Yamasaki}, {Totani}, \&
  {Kiuchi}}]{2017arXiv171002302Y}
{Yamasaki}, S., {Totani}, T., \& {Kiuchi}, K. 2017, ArXiv e-prints,
  arXiv:1710.02302

\bibitem[{{Yang} \& {Zhang}(2016)}]{2016ApJ...830L..31Y}
{Yang}, Y.-P., \& {Zhang}, B. 2016, \apjl, 830, L31

\bibitem[{{Zhang}(2016)}]{2016ApJ...827L..31Z}
{Zhang}, B. 2016, \apjl, 827, L31

\bibitem[{{Zhao} {et~al.}(2011){Zhao}, {van den Broeck}, {Baskaran}, \&
  {Li}}]{2011PhRvD..83b3005Z}
{Zhao}, W., {van den Broeck}, C., {Baskaran}, D., \& {Li}, T.~G.~F. 2011, \prd,
  83, 023005

\bibitem[{{Zheng} {et~al.}(2014){Zheng}, {Ofek}, {Kulkarni}, {Neill}, \&
  {Juric}}]{2014ApJ...797...71Z}
{Zheng}, Z., {Ofek}, E.~O., {Kulkarni}, S.~R., {Neill}, J.~D., \& {Juric}, M.
  2014, \apj, 797, 71

\bibitem[{{Zhou} {et~al.}(2014){Zhou}, {Li}, {Wang}, {Fan}, \&
  {Wei}}]{2014PhRvD..89j7303Z}
{Zhou}, B., {Li}, X., {Wang}, T., {Fan}, Y.-Z., \& {Wei}, D.-M. 2014, \prd, 89,
  107303

\end{thebibliography}

\end{document}